\documentstyle[12pt]{article}

\newcommand{\begeq}{\begin{equation}}
\newcommand{\enqu}{\end{equation}}
\newcommand{\begar}{\begin{array}}
\newcommand{\enay}{\end{array}}

\newcommand{\lm}{\lambda}
\newcommand{\tea}{\tau}
\newcommand{\al}{\alpha}
\newcommand{\be}{\beta}
\newcommand{\Al}{{\cal M}}

\newcommand{\ds}{\displaystyle}
\begin{document}

\begin{center}
{\Large\bf Fine Structure of Matrix Darboux-Toda Integrable Mapping}

\vspace{1cm}

A.~N.~Leznov$^a$ and E.~A.~Yuzbashyan$^b$

\vspace{1cm}

{\small\it
\noindent $^a$Institute for High Energy Physics,
142284 Protvino, Moscow Region, Russia.
 
\noindent $^b$Bogoliubov Laboratory of Theoretical Physics,
Joint Institute for Nuclear Research,\\
141980 Dubna, Moscow Region, Russia.
 
}

\end{center}

\begin{abstract}
 We show here that matrix Darboux-Toda transformation can be written as a product of a number
of mappings. Each of these mappings is a symmetry of the matrix nonlinear Shr\"odinger system
of integro--differential equations. We thus introduce a completely new type of discrete
transformations for this system. The discrete symmetry of the vector nonlinear
Shr\"odinger system is a particular realization  of these
mappings.
\end{abstract}
\begin{quote}
 \end{quote}

\section{Introduction}

 Discrete transformations that leave an integrable system invariant (integrable mappings) provide an important tool in the theory of integrable systems. Furthermore, it has been suggested  \cite{1}
that  the theory of integrable systems is closely related to a representation theory of the group of integrable mappings.  In this approach a classification of  integrable mappings plays a key role.

Recently, a discrete symmetry of the vector nonlinear Shr\"odinger system  (the $v$-mapping in
future references) was discovered by Aratyn \cite{3}. To find it, the author considered transformations that preserve the form of the corresponding Lax operator and equation (a technique that can be applied in the (1+1)d case only).

In the present paper we reveal new discrete symmetries of the (1+2)d matrix
nonlinear Shr\"odinger  system (MNLSS) \cite{4, Nucl} and show that the $v$-mapping (generalized
to two spacial dimensions) is a particular case of these symmetries.

The paper is organized as follows.  In Section 2 we write down soliton-like solutions of (1+1)d MNLSS. As far as we are aware this form of solutions was not previously known. In Section 3 we show
that these solutions are invariant under a certain class of discrete transformations. Moreover, we
find that these  symmetries are not limited to a particular type of solutions of MNLSS, i.e.  these are
symmetries of MNLSS itself. These results are generalized to (1+2)d case in Section 4. Finally,
in Section  5  we discuss the relationship between different discrete symmetries of MNLSS and show that the $v$-mapping  is a particular case of discrete transformations introduced in the present paper.

\section{Multi--Soliton Solutions}
 
Here we write down explicit expressions for multi-soliton type solutions of
(1+1)d MNLSS.  Details and proofs can be found in \cite{Nucl}. 

MNLSS in (1+1)d is the following  system of two coupled nonlinear differential equations:
\begeq
\begar{r}
-v_t+v_{xx}+2vuv=0\\
u_t+u_{xx}+2uvu=0\\
\enay \label{NS}
\enqu
where $u$ and $v$ are $k\times k$ matrix functions of $t$ and $x$.
In particular, when the non-zero part of $v$ is
a single column and that of $u$ is a single row,  system (\ref{NS}) reduces to 
the vector nonlinear Shr\"odinger system  (VNLSS).

 Multi-soliton type solutions
of MNLSS (\ref{NS}) can be characterized  by a pair of vectors $(\vec n, \vec m)$ with 
integer valued components $n_i$ and
$m_j$ such that  ${n_i\ge -1}$ and ${m_j\ge -1}$. The solutions can be written symbolically as
\begeq
\begar{l}
\ds u_{ij}=-{|n_1,\dots,n_i+1,\dots,n_k;m_1,\dots,m_j-1,\dots,m_k|\over
|n_1,\dots,n_k;m_1,\dots,m_k|}\\
\\
\ds v_{ij}={|n_1,\dots,n_j-1,\dots,n_k;m_1,\dots,m_i+1,\dots,m_k|\over
|n_1,\dots,n_k;m_1,\dots,m_k|}\\
\enay\label{SS}
\enqu
where ${|n_1,\dots,n_k;m_1,\dots,m_k|}$ is the determinant of
a matrix whose rows are split into segments of lengths
${n_1+1,\dots,n_k+1;m_1+1,\dots,m_k+1}$.  Each segment is represented by an integer
in equations (\ref{SS}). Segments of
the $s$th row that correspond to integers $n_i$ and $m_j$ are
$$
\begar{l}
\ds n_i:\qquad e^{2\tea_s}, e^{2\tea_s}\lm_s,\dots, e^{2\tea_s}\lm_s^{n_i}\\
\\
\ds m_j:\qquad 1,\lm_s,\dots,\lm_s^{m_j}\\
\enay
$$
where $\tea_s=\lm_s x/2-\lm_s^2 t/4+c_s$, and $\lm_s$ and $c_s$ are 
arbitrary parameters.
For example, 
$$
|0;1|=
\begar{|ccc|}
e^{2\tea_1}&1&\lm_1\\
e^{2\tea_2}&1&\lm_2\\
e^{2\tea_3}&1&\lm_3\\
\enay\qquad
|1;-1|=
\begar{|cc|}
e^{2\tea_1}&e^{2\tea_1}\lm_1\\
e^{2\tea_2}&e^{2\tea_2}\lm_2\\
\enay
$$

Using  identity (\ref{d}) from  Appendix, one can check that $u$ and $v$ given by (\ref{SS}) are indeed solutions of
MNLSS (\ref{NS}). Solutions of VNLSS
are obtained from (\ref{SS}) by setting ${m_j=-1}$ for ${j\ge 2}$.

\section{Discrete Symmetries of MNLSS}

In this section we derive mappings between different  multi-soliton type solutions (\ref{SS}).
Furthermore, we show that these mappings transform any solution of (1+1)d MNLSS (\ref{NS})
into another solution of the same system, i.e. these are symmetries of MNLSS.

First, let us replace a pair of integers in (\ref{SS}),  say   $n_\al$ and $n_\be$, with ${n_{\al}-1}$ and ${m_{\be}+1}$ respectively and denote the resulting solution by ($\tilde u$, $\tilde v$). 
It follows from (\ref{SS})  that
$$
\tilde u_{\al\be}={1\over v_{\be\al}}
$$
Using identity (\ref{d}) from Appendix, one can also establish the following
relations between solutions ($u, v$) and ($\tilde u, \tilde v$):
\begeq
\begar{l} \ds (\tilde u_{i\be} v_{\be\al})_x=-(uv)_{i\al}\quad \ds (\tilde
u_{\al j} v_{\be\al})_x=-(vu)_{\be j}\quad \ds \tilde u_{\al\be}={1\over
v_{\be\al}}\\\\

\ds \left({v_{\be i}\over v_{\be\al}}\right)_x=(\widetilde{uv})_{\al i}\quad
\ds \left({v_{j\al}\over v_{\be\al}}\right)_x=(\widetilde{vu})_{j\be}\\\\

\ds \tilde v_{ji}=v_{ji}-{v_{\be i}v_{j\al}\over v_{\be\al}}\quad
\ds \tilde u_{ij}=u_{ij}+\tilde u_{i\be}\tilde u_{\al j} v_{\be\al}\\

\ds v^2_{\be\al}(\widetilde{uvu})_{\al\be}-(vuv)_{\be\al}=
v_{\be\al}(\ln v_{\be\al})_{xx}\\
\enay\label{s}
\enqu
where $i\ne\al$ and $j\ne\be$.  Note that since $\alpha$ and $\beta$ can take $k$ values each, there are $k^2$ basic 
mappings.

Relations (\ref{s}) connect  different  solutions of  MNLSS
(\ref{NS}) of a particular type (multi-soliton). However, it turns out that mappings (\ref{s})  are not limited to
this type of solutions. Indeed,  by a direct substitution of (\ref{s}) into
(\ref{NS}) one can check that MNLSS (\ref{NS})  is invariant under transformations (\ref{s}).
A product of any number of mappings
(\ref{s}) is clearly also a discrete symmetry of MNLSS (\ref{NS}).

\section{2d Case}

Here we generalize the result of the previous section (equations (\ref{s})) to the case of two spatial dimensions.
 
In this case,  MNLSS \cite{4, Nucl} (also called 2d matrix Davey--Stewartson system) reads
\begeq
\left\{
\begin{array}{r}
\ds -v_t+a v_{xx}+b v_{yy}+2a\int dy (vu)_x v+2b v\int dx (uv)_y =0\\
\ds  u_t+a u_{xx}+b u_{yy}+2au\int dy (vu)_x+2b\int dx (uv)_y u=0\\
\end{array}\right.\label{ns2}
\enqu
where $a$ and $b$ are arbitrary numbers, and $u$ and $v$ are $k\times k$ matrix functions
of $t$, $x$, and $y$ .  Note that by setting $x=y$,  appropriately choosing constants of integration,
and rescaling the time variable $t$, we can reduce system (\ref{ns2}) to its
1d counterpart (\ref{NS}).

System (\ref{ns2}) is
the third member of the (1+2)d matrix nonlinear Shr\"odinger hierarchy \cite{5} of integrable systems.
This hierarchy is an infinite set of integrable (1+2)d matrix  nonlinear integro-differential equations
all of which are invariant under the following transformation (matrix Darboux--Toda mapping):
\begin{equation}
\tilde u=v^{-1}\quad \tilde v=[vu-(v_x v^{-1})_y]
v\equiv v [uv-(v^{-1}v_y)_x]\label{m}
\end{equation}
where $u$ and $v$ are assumed to be nonsingular. 
 Mapping (\ref{s}) can be generalized to the case of two spatial dimensions as follows
\begeq \begar{l} \ds
(\tilde u_{i\be} v_{\be\al})_x=-(uv)_{i\al}\quad \ds (\tilde u_{\al j}
v_{\be\al})_y=-(vu)_{\be j}\quad \ds \tilde u_{\al\be}={1\over v_{\be\al}}\\\\

\ds \left({v_{\be i}\over v_{\be\al}}\right)_x=(\widetilde{uv})_{\al i}\quad
\ds \left({v_{j\al}\over v_{\be\al}}\right)_y=(\widetilde{vu})_{j\be}\\\\

\ds \tilde v_{ji}=v_{ji}-{v_{\be i}v_{j\al}\over v_{\be\al}}\quad
\ds \tilde u_{ij}=u_{ij}+\tilde u_{i\be}\tilde u_{\al j} v_{\be\al}\\

\ds v^2_{\be\al}(\widetilde{uvu})_{\al\be}-(vuv)_{\be\al}=
v_{\be\al}(\ln v_{\be\al})_{xy}\\
\enay\label{s2}
\enqu
 Plugging $\tilde u$ and $\tilde v$ instead of $u$ and $v$ into (\ref{ns2}) and using 
equations (\ref{s2}), one can verify  that equations (\ref{ns2}) are invariant under 
mapping (\ref{s2}).

In the present paper we do not address the problem of constructing a hierarchy
 of equations invariant under only a single mapping  (\ref{s2}) with a particular choice of
$\alpha$ and $\beta$.  However, one can show that
all equations of (1+2)d matrix nonlinear Shr\"odinger hierarchy are invariant under all transformations (\ref{s2}). This follows from the fact that matrix Darboux--Toda mapping (\ref{m})
commutes with all mappings (\ref{s2})  (see Section 5) and from the construction of the hierarchy
\cite{5}.

\section{Relations between symmetries of MNLSS}

In this section we discuss the relationship between different symmetries of MNLSS, in particular  between the $v$-mapping of Ref. \cite{3}, matrix Darboux--Toda mapping (\ref{m}), and transformations (\ref{s}) and (\ref{s2}) derived in the present paper.

First, the mapping for (1+1)d  vector nonlinear Shr\"odinger system of Ref. \cite{3} can be obtained from symmetries (\ref{s}) by setting
$$
\begar{l}
\ds \al=\be=r\qquad u_{ir}\equiv u_i\qquad v_{ri}\equiv v_i\\
\ds u_{ij}=v_{ji}=0\qquad j\ne r\\
\enay
$$
i.e. the $v$-mapping is a particular case of these symmetries.

Next, let us consider  matrix Darboux--Toda mapping (\ref{m}) and transformations  (\ref{s2}).
Let $T_{\alpha\beta}$ and $M_k$ denote mappings (\ref{s2}) and (\ref{m}) respectively. One can
prove the following relations:  
$$
T_{11} T_{22}\dots T_{kk}=M_k
$$
$$
T_{ij}T_{ji}=T_{ii}T_{jj}\qquad T_{ij}T_{jk}T_{ki}=T_{ii}T_{jj}T_{kk}\qquad\dots
$$
$$
T_{ij}T_{i'j'}=T_{i'j'}T_{ij}\qquad T_{ij} M_k=M_k T_{ij}
$$

We conclude this section by noting that in (1+1)d case MNLSS possesses an additional discrete symmetry
\begin{equation}
\tilde u_x=u-\tilde u v\tilde u\qquad v_x=v\tilde u v-\tilde v\label{another}
\end{equation}
We  checked that when $u$ and $v$ are scalar valued functions,
this mapping and the 1d counterpart of Darboux-Toda transformation (obtained by setting $x=y$ in (\ref{m}))  produce the same solutions of (1+1)d nonlinear Shr\"odinger system.  However, presently
it is not clear to us whether  symmetry (\ref{another}) has a 2d analogue.

\section{Conclusion}

We found new discrete symmetries (see equations (\ref{s})) of (1+2)d matrix nonlinear Shr\"odinger system. While  Darboux-Toda transformation (\ref{m}) is limited to nonsingular square
matricies, mappings (\ref{s}) are free from this constraint. In particular, when matricies $u$
and $v$ in equations (\ref{ns2}) and (\ref{s2}) reduce to a single row and column respectively, we obtain (1+2)d generalizations of the vector nonlinear Shr\"odinger system and
the corresponding symmetry.

Presently, we do not know how to construct hierarchies of integrable systems that are invariant
only under one of mappings (\ref{s2}). The conventional
Lax technique is not applicable to the 2d case and an approach similar to that employed in \cite{5}
might be required. 
   
\section{Appendix}

Here we derive an identity
\begeq
|{\cal M} C_1 C_2| |\Al C_3 C_4| + |\Al C_2 C_3| |\Al C_1 C_4|= |\Al C_2 C_4| |\Al C_1 C_3|\label{d}
\enqu
where $\Al$ is an arbitrary $k\times(k-2)$ matrix, 
$C_1, C_2, C_3$, and $C_4$ are columns of lengths $k$, and $|~|$ denotes the determinant.

First, consider a $k\times k$  matrix
$$
F=
\left(\begar{ccc}
A & a_1 & b_1\\
a_2 & c_1 & d_1\\
b_2 & c_2 & d_2\\
\enay\right)
$$
where $A$ is a $(k-2)\times (k-2)$ matrix,  $a_1$ and  $b_1$, and $a_2$ and $b_2$ are columns and
rows respectively of length $k$ each,  and $c_{1,2}$ and $d_{1,2}$ are
 scalars. 
It is simple to show that
$$
\begar{l}
|F|=|A|\
 \left|\begar{ccc}
     E & A^{-1}a_1 & A^{-1}b_1\\
     a_2 & c_1 & d_1\\
     b_2 & c_2 & d_2\\
     \enay \right|
=|A|\  \left| \begar{cc}
                       c_1-a_2A^{-1}a_1 & d_1-a_2A^{-1}b_1 \\
                       c_2-b_2A^{-1}a_1 & d_2-b_2A^{-1}a_1 \\
                       \enay \right|
\enay
$$
where $E$ is the $k\times k$ unit matrix.  Now identity (\ref{d}) can be proven by applying the above equation to each term in equation (\ref{d}).

\end{document}